\documentclass[aip, jcp, superscriptaddress, reprint, onecolumn]{revtex4-1}
\usepackage[utf8]{inputenc}
\usepackage{bm}
\usepackage{amsmath}
\usepackage{amsfonts}
\usepackage{amssymb}
\usepackage{amsthm}
\usepackage{xcolor}
\usepackage[breaklinks=true,colorlinks=true,linkcolor=blue,urlcolor=blue,citecolor=blue]{hyperref}
\usepackage{graphicx}

\newcommand{\newcite}[1]{\textcolor{violet}{\cite{#1}}}

\newcommand{\defeq}{\mathrel{\mathop:}=}

\newcommand{\U}{\mathcal{U}}
\newcommand{\D}{\mathcal{D}}
\renewcommand{\H}{\mathcal{H}}
\newcommand{\params}{\boldsymbol{\lambda}}

\begin{document}

\title{Kinetic energy fluctuations and specific heat in generalized ensembles}

\author{Sergio Davis}
\email{sergio.davis@cchen.cl}

\affiliation{Research Center in the intersection of Plasma Physics, Matter and Complexity (P$^2$mc), Comisión Chilena de Energía Nuclear, Casilla 188-D, Santiago, Chile}
\affiliation{Departamento de Física y Astronomía, Facultad de Ciencias Exactas, Universidad Andres Bello. Sazié 2212, piso 7, 8370136, Santiago, Chile.}

\author{Catalina Ru\'iz}
\affiliation{Departamento de Física y Astronomía, Facultad de Ciencias Exactas, Universidad Andres Bello. Sazié 2212, piso 7, 8370136, Santiago, Chile.}

\author{Claudia Loyola}
\affiliation{Departamento de Física y Astronomía, Facultad de Ciencias Exactas, Universidad Andres Bello. Sazié 2212, piso 7, 8370136, Santiago, Chile.}

\author{Carlos Femen\'ias}
\affiliation{Departamento de Física y Astronomía, Facultad de Ciencias Exactas, Universidad Andres Bello. Sazié 2212, piso 7, 8370136, Santiago, Chile.}

\author{Joaqu\'in Peralta}
\email{joaquin.peralta@unab.cl}
\affiliation{Departamento de Física y Astronomía, Facultad de Ciencias Exactas, Universidad Andres Bello. Sazié 2212, piso 7, 8370136, Santiago, Chile.}

\begin{abstract}
We derive an exact generalization of the well-known Lebowitz--Percus--Verlet (LPV) formula that relates the kinetic energy fluctuations of an isolated system to its specific heat. Our general formula, obtained by the 
application of expectation identities, is valid for arbitrary steady--state ensembles and system sizes, expressing the relative variance of the kinetic energy in terms of the variance of total 
energy and the microcanonical specific heat. The usual microcanonical LPV formula can be readily recovered as a particular case where energy fluctuations vanish. We test the validity of the generalized formula by performing 
Monte Carlo simulations of a superstatistical system of harmonic oscillators, as well as by exact calculation of energy variances in a uniform--energy ensemble, discussing its relevance to systems exhibiting negative heat 
capacity and ensemble inequivalence, as encountered in finite nuclei and self--gravitating models. Our results may provide useful in the study of non-equilibrium phase transitions in finite systems.
\end{abstract}

\maketitle

\section{Introduction}
Studying fluctuations in physical observables within statistical mechanics is crucial for understanding the distinct characteristics of various statistical ensembles~\cite{pathria2016, huang1987statistical}. Among these ensembles, the microcanonical ensemble holds a fundamental place~\cite{hill1986introduction, gross2001microcanonical}. This ensemble, where energy, volume, and the number of particles are conserved, allows for a focused examination of isolated systems. By analyzing the fluctuations of physical observables within the microcanonical framework, we can clearly depict the underlying behaviors of these systems. One important observable is the specific heat ($C$), typically measured at constant volume. The specific heat is a crucial quantity that provides insights into a wide range of material properties, particularly because of its unique behavior near phase transitions~\cite{Schierz2016, stanley1971introduction}.

Beyond the thermodynamic limit, the behavior of finite and isolated systems reveals a richer phenomenology. In particular, negative heat capacities and ensemble inequivalence have been observed in a variety of finite systems, ranging from fragmenting nuclei and atomic clusters to long--range self--gravitating models\newcite{Borderie2022,Maciel2017,Przedborski2017,LyndenBell1998,Ota2001}. These anomalies emerge near first--order phase transitions and signal the presence convex intruders in the microcanonical entropy, implying that the canonical and microcanonical ensembles can yield different equilibrium behaviors\newcite{Borderie2022,Maciel2017}. Partial energy fluctuations have been proposed as a diagnostic of such phase transitions and have been used to reconstruct the microcanonical heat capacity even in small systems\newcite{Borderie2022}. Understanding kinetic energy fluctuations in generalized ensembles may provide a diagnostic tool that could, in principle, be applied to systems known to exhibit negative heat capacities and ensemble inequivalence.

A key theoretical link between these anomalous behaviors and equilibrium thermodynamic quantities is provided by the relationship between kinetic energy fluctuations and the specific heat. A landmark result by Lebowitz, Percus, and Verlet (LPV)~\cite{Lebowitz1967} established that the microcanonical fluctuations of kinetic energy are related to the specific heat. This relationship provides a deeper understanding of how specific heat can describe the fluctuations and behavior within different statistical ensembles. The formula is given by
\begin{equation}
\label{eq:LPV}
\big<(\delta K)^2\big>_E = \frac{3N}{2{\beta_E}^2}\left(1-\frac{3N}{2C}\right),
\end{equation}
where $\big<(\delta K)^2\big>_E$ is the variance of kinetic energy, $N$ is the number of particles, 
\begin{equation}
\beta_E \defeq \frac{1}{k_B T(E)}
\end{equation}
is the inverse temperature at energy $E$, and $C$ is the total specific heat of the system in units of $k_B$.

Studies have shown that partial energy fluctuations can be used to reconstruct microcanonical heat capacities even in small systems and near phase transitions. For example, in the context of nuclear multrifragmentation, the relative variance of the kinetic energy, $\sigma_K^2/T^2$, provides an estimate of the heat capacity; simulations with lattice gas and Lennard--Jones models validate this approach\cite{Gulminelli2005}. These results motivate our present generalization of the LPV formula and emphasize the importance of kinetic energy fluctuations in detecting phase--transition signatures.

Small and finite systems also exhibit fluctuations of intensive quantities such as temperature, which cannot be neglected. The formalism of superstatistics~\cite{Beck2003, Beck2004} has been advocated as a framework for 
describing temperature fluctuations in nonequilibrium steady states; however, its interpretation of $\beta$ as a fluctuating inverse temperature faces conceptual difficulties~\cite{Davis2021ft}. Recent work has introduced 
generalized ensembles described by an ensemble function $\rho(E; \params)$ and shown that superstatistics describes only a subclass of such states. Particularly, the inverse temperature covariance
\begin{equation}
\label{eq:U}
\U \defeq \big<(\delta \beta_\Omega)^2\big>_{\params} + \big<{\beta_\Omega}'\big>_{\params},
\end{equation}
where
\begin{equation}
\beta_\Omega(E) \defeq \frac{\partial}{\partial E}\ln \Omega(E)
\end{equation}
is the microcanonical inverse temperature and $\Omega(E)$ is the density of states, allows to classify nonequilibrium steady states into supercanonical states (including superstatistics), where $\U > 0$ and 
subcanonical states~\cite{Davis2022b}, with $\U < 0$. This classification helps to identify whether temperature fluctuations can be encoded by a distribution $P(\beta|\params)$ or more general descriptions are needed.

In this study, we propose to investigate the fluctuations of kinetic energy and their relationship to the specific heat, aiming to extend our understanding to different ensembles, in this particular case, the generalized ensemble~\cite{costeniuc2006generalized, mitsutake2001generalized, Moreno2022}. From here, we derive a generalization of the LPV formula
for an arbitrary steady-state ensemble with parameters $\params$, namely
\begin{equation}
\label{eq:main}
\frac{\big<(\delta K)^2\big>_{\params}}{\big<K\big>_{\params}^2} =
\frac{2}{3N}\frac{C+1}{C+2}\left[\left(\frac{3N}{2}+1\right)\frac{\big<(\delta E)^2\big>_{\params}}{\big<E\big>_{\params}^2} + 1-\frac{3N}{2(C+1)}\right],
\end{equation}
where now $\big<(\delta E)^2\big>_{\params}$ represents the energy fluctuations within the ensemble parameterized by $\params$, while $\big<E\big>_{\params}$ is its mean total energy. Similarly, $\big<K\big>_{\params}$ 
stands for the mean kinetic energy in the ensemble. Importantly, our method for arriving at this result does not rely on the assumptions used in the original LPV formula, and it holds true for any system size $N \geq 1$.

In the following sections, we will outline the structure of the paper. Section 2 will detail the derivation of the generalized equation~\ref{eq:main}, and Section 3 will explain our computational procedure using Monte Carlo simulations to test our findings. We will conclude with some final remarks in Section 4.
\section{Derivation of the main result}
\label{sec:derivation}

\noindent
We will assume a classical system of $N$ particles with Hamiltonian $\mathcal{H}$ given by
\begin{equation}
\mathcal{H}(\bm{r}_1, \ldots, \bm{r}_N, \bm{p}_1, \ldots, \bm{p}_N) = \sum_{i=1}^N \frac{\bm{p}_i^2}{2m_i} + \Phi(\bm{r}_1,\ldots,\bm{r}_N)
\end{equation}
and constant specific heat $C$ in units of $k_B$, that is, such that
\begin{equation}
\label{eq:betaomega_C}
\beta_\Omega(E) = \frac{C}{E},
\end{equation}
with $\Omega(E)$ the density of states,
\begin{equation}
\Omega(E) \defeq \int d\bm{r}_1\ldots d\bm{r}_N d\bm{p}_1 \ldots d\bm{p}_N \delta\left(E- \left[\sum_{i=1}^N \frac{\bm{p}_i^2}{2m_i} + \Phi(\bm{r}_1,\ldots,\bm{r}_N)\right]\right),
\end{equation}
such that, in our case, it can be obtained from \eqref{eq:betaomega_C} as $\Omega(E) = \Omega_0\,E^C$.

We will assume that the interaction potential $\Phi$ is bounded from below, and without loss of generality, we will set this minimum value to zero so that the energy is always non-negative.

From here, we will consider the system in a non-equilibrium, steady-state ensemble described by the microstate distribution:
\begin{equation}
P(\bm R, \bm P|\params) = \rho(\H(\bm R, \bm P); \params)
\end{equation}
where $\bm{R} \defeq (\bm{r}_1, \ldots, \bm{r}_N)$ and $\bm{P} \defeq (\bm{p}_1, \ldots, \bm{p}_N)$, and the function $\rho(E; \params)$ is the \emph{ensemble function}.
The distribution of total energy is simply
\begin{equation}
P(E|\params) = \rho(E; \params)\Omega(E),
\end{equation}
while the distribution of kinetic energy in this ensemble can be computed from
\begin{equation}
\label{eq:kindist}
P(K|\params) = \int d\bm{R}d\bm{P} \;\rho\big(K(\bm P) + \Phi(\bm R); \params\big)\delta\big(K(\bm P)-K\big) = \left[\int d\bm{R}\; \rho(K+\Phi(\bm R); \params)\right]\;\Omega_K(K; N)
\end{equation}
where $\Omega_K$ is the kinetic density of states, which is of the form
\begin{equation}
\label{eq:omegaK}
\Omega_K(K; N) = W_N\;K^{\frac{3N}{2}-1},
\end{equation}
with $W_N$ a constant only dependent on $N$ and the particle masses. We can replace the multidimensional integral in square brackets with a one-dimensional integral weighted by the configurational density of 
states $\D(\phi)$,
\begin{equation}
\int d\bm{R} \;\rho(K+\Phi(\bm R); \params) = \int_0^\infty d\phi \D(\phi)\;\rho(K+\phi; \params),
\end{equation}
where
\begin{equation}
\D(\phi) \defeq \int d\bm{R}\delta(\Phi(\bm R)-\phi).
\end{equation}

\noindent
Replacing \eqref{eq:omegaK} we obtain
\begin{equation}
P(K|\params) = W_N K^{\frac{3N}{2}-1}\int_0^\infty d\phi \D(\phi)\rho(K+\phi; \params).
\end{equation}

\noindent
Moreover, the full density of states $\Omega(E)$ can be computed from $\D(\phi)$ as
\begin{equation}
\Omega(E) = W_N\,\int_0^\infty dK\,K^{\frac{3N}{2}-1}\D(E-K).
\end{equation}

\noindent
In the case of the microcanonical ensemble, we have
\begin{equation}
\rho(K+\phi; E) = \frac{\delta(K+\phi-E)}{\Omega_0 E^C},
\end{equation}
therefore the microcanonical distribution of kinetic energies is
\begin{equation}
\label{eq:pke_free}
P(K|E) = \frac{W_N}{\Omega_0}\frac{K^{\frac{3N}{2}-1}\D(E-K)}{E^C}.
\end{equation}

In the following, we will compute the first and second moments of this kinetic energy distribution given $E$, without explicit knowledge of the configurational density of states $\D$, by virtue of 
two expectation identities, namely the conjugate variables theorem (CVT)~\cite{Davis2012} and the fluctuation-dissipation theorem (FDT)~\cite{Davis2016c}. For $P(K|E)$ in \eqref{eq:pke_free}, the CVT takes the form
\begin{equation}
\big<g'\big>_E = \left<g\left[\frac{\partial}{\partial K}\ln \D(E-K)-\frac{3N-2}{2K}\right]\right>_E
\end{equation}
where $g = g(K; E)$ is an arbitrary function of $K$ and $E$ such that $g' \defeq \partial g/\partial K$ exists. On the other hand, the FDT for the same distribution is
\begin{equation}
\frac{\partial}{\partial E}\big<g\big>_E = -\left<g\,\frac{\partial}{\partial K}\ln \D(E-K)\right>_E -\frac{C}{E}\big<g\big>_E.
\end{equation}

\noindent
Combining the two we can eliminate the term depending on $\D$ to obtain
\begin{equation}
\label{eq:mega}
\frac{\partial}{\partial E}\big<g\big>_E = \big<g'\big>_E + \Big(\frac{3N}{2}-1\Big)\left<\frac{g}{K}\right>_E - \frac{C}{E}\big<g\big>_E.
\end{equation}

\noindent
Using $g(K; E) = K$ and $g(K; E) = K^2$ in \eqref{eq:mega} we can solve for the first and second moment of $P(K|E)$, obtaining
\begin{subequations}
\begin{align}
\label{eq:ke}
\big<K\big>_E & = \frac{3NE}{2(C+1)}, \\
\label{eq:k2e}
\big<K^2\big>_E & = \frac{3N(3N+2)E^2}{4(C+1)(C+2)},
\end{align}
\end{subequations}
respectively. Taking the expectation of \eqref{eq:ke} and \eqref{eq:k2e} under given $\params$, we have that the first and second moments of the kinetic energy in the generalized ensemble are 
proportional to the corresponding moments of the total energy distribution given $\params$. More precisely,
\begin{subequations}
\begin{align}
\big<K\big>_{\params} & = \frac{3N}{2(C+1)}\big<E\big>_{\params}, \\
\label{eq:k2s}
\big<K^2\big>_{\params} & = \frac{3N(3N+2)}{4(C+1)(C+2)}\big<E^2\big>_{\params},
\end{align}
\end{subequations}
respectively, from which we can obtain the relationship between the variances of $K$ and $E$ in the ensemble parameterized by $\params$ as
\begin{equation}
\big<(\delta K)^2\big>_{\params} = \frac{3N(3N+2)}{4(C+1)(C+2)}\Big[\big<(\delta E)^2\big>_{\params} + \big<E\big>_{\params}^2\Big] - \left(\frac{3N}{2(C+1)}\right)^2\big<E\big>_{\params}^2.
\end{equation}

By combining the terms, dividing by $\big<K\big>_{\params}^2$ according to \eqref{eq:k2s} and after some algebra, we obtain \eqref{eq:main}, our main result. Our result is valid for finite systems, as it does not 
assume the thermodynamic limit. Interestingly, it is also valid for negative heat capacities, provided that $C > -1$. We can easily recover a finite-size version of the original LPV in \eqref{eq:LPV}, for this we 
simply set the variance of $E$ to zero, replace $\big<K\big>_E$ using \eqref{eq:ke} and obtain
\begin{equation}
\big<(\delta K)^2\big>_E = \frac{3N E^2}{2(C+1)(C+2)}\left(1-\frac{3N}{2(C+1)}\right),
\end{equation}
which reduces to \eqref{eq:LPV} when $C + 2 \approx C + 1 \approx C$, by recognizing $\beta_E = \beta_\Omega(E) = C/E$. On the other hand, taking the limit $N \rightarrow \infty$ 
of \eqref{eq:main} we see that it reduces to the equality of the relative variances of $K$ and $E$,
\begin{equation}
\frac{\big<(\delta K)^2\big>_{\params}}{\big<K\big>_{\params}^2} = \frac{\big<(\delta E)^2\big>_{\params}}{\big<E\big>_{\params}^2}.
\end{equation}

It is also clear that the fluctuations of the total energy contribute to increasing fluctuations of kinetic energy, as expected and in agreement with the law of total variance~\cite{Weiss2006},
\begin{equation}
\big<(\delta K)^2\big>_{\params} = \Big<\big<(\delta K)^2\big>_E\Big>_{\params} + \big<(\delta \mathcal{K}_E)^2\big>_{\params}
\end{equation}
with $\mathcal{K}_E \defeq \big<K\big>_E$. 

\section{Verification}

\subsection{Canonical ensemble}

In the case of the canonical ensemble, we know that for $\Omega(E) \propto E^C$ the full partition function is
\begin{equation}
Z(\beta) = \int_0^\infty dE\,\Omega(E)\exp(-\beta E) \propto \beta^{-(C+1)}
\end{equation}
therefore
\begin{equation}
\big<E\big>_\beta = -\frac{\partial}{\partial \beta}\ln Z(\beta) = \frac{C+1}{\beta}.
\end{equation}

\noindent
On the other hand, from $\Omega_K(K)$ we obtain the kinetic part of the partition function, namely
\begin{equation}
Z_K(\beta) = \int_0^\infty dE\,\Omega_K(K)\exp(-\beta K) \propto \beta^{-\frac{3N}{2}}
\end{equation}
thus
\begin{equation}
\big<K\big>_\beta = -\frac{\partial}{\partial \beta}\ln Z_K(\beta) = \frac{3N}{2\beta}.
\end{equation}
and from them we have
\begin{equation}
\label{eq:varE_canon}
\big<(\delta E)^2\big>_\beta = -\frac{\partial \big<E\big>_\beta}{\partial \beta} = \frac{C+1}{\beta^2},
\end{equation}
and
\begin{equation}
\big<(\delta K)^2\big>_\beta = -\frac{\partial \big<K\big>_\beta}{\partial \beta} = \frac{3N}{2\beta^2}.
\end{equation}

\noindent
In this way, the relative variances of $E$ and $K$ are given by
\begin{equation}
\label{eq:relvarE_canon}
\frac{\big<(\delta E)^2\big>_\beta}{\big<E\big>_{\beta}^2} = \frac{1}{C+1},
\end{equation}
and
\begin{equation}
\label{eq:relvarK_canon}
\frac{\big<(\delta K)^2\big>_\beta}{\big<K\big>_{\beta}^2} = \frac{2}{3N},
\end{equation}
respectively. Replacing \eqref{eq:relvarE_canon} and \eqref{eq:relvarK_canon} into \eqref{eq:main} and after some algebra, we see that \eqref{eq:main} holds exactly for all $N \geq 1$.

\subsection{Superstatistical harmonic oscillator ensemble}
\label{sec:ho-superstat}

In order to illustrate Eq.~\eqref{eq:main} beyond the conventional canonical ensemble, we consider a simple non--equilibrium model in which the system samples a distribution of temperatures.  Following the framework of 
superstatistics, a driven or inhomogeneous system can be regarded as a mixture of canonical ensembles with a fluctuating inverse temperature $\beta$ drawn from a probability density 
$P(\beta|\params)$. Specifically, we consider a set of $N$ identical particles of mass $m$ moving in three dimensions under an isotropic harmonic potential $V(\bm{r})=\tfrac{1}{2} \omega^2\,\|\bm{r}\|^2$. The 
Hamiltonian of this system reads
\begin{equation}
\mathcal{H}(\bm{r}_1, \ldots, \bm{r}_N, \bm{p}_1, \ldots, \bm{p}_N) = \sum_{i=1}^N \left[\frac{\bm{p}_i^2}{2m} + \frac{1}{2} m\omega^2\,\bm{r}_i^2 \right],
\end{equation}
where $\bm{r}_i$ and $\bm{p}_i$ denote the position and momentum of particle $i$, respectively.  Because the energy is a sum of quadratic terms, the microcanonical density of states scales as $\Omega(E)\propto E^C$ 
with $C = 3N-1$, which is precisely the constant specific heat in units of $k_B$. We introduce temperature fluctuations by assuming that the inverse temperature $\beta$ follows a gamma distribution with $\alpha$ and 
$\theta$ known as the shape and scale parameter, so we have $\params = (\alpha, \theta)$ with
\begin{equation}
P(\beta|\alpha, \theta) = \frac{1}{\Gamma(\alpha) \theta^\alpha} \beta^{\alpha - 1} \exp\left(-\frac{\beta}{\theta}\right).
\end{equation}

This choice of distribution for $\beta$ is such that the superstatistical ensemble of the system corresponds to a $q$-canonical ensemble, sometimes referred to as Tsallis statistics~\cite{Tsallis1988, Tsallis2009}.

For each canonical configuration used in our superstatistics approach, we perform Monte Carlo sampling by drawing $\bm{p}_i$ and $\bm{r}_i$ independently from Gaussian distributions with variances 
$\langle p_{i,\alpha}^2 \rangle = m/\beta$ and $\langle r_{i,\alpha}^2 \rangle = 1/(m\omega^2\beta)$ ($\alpha$ labels Cartesian components). For each one of the $\beta$, we compute the kinetic energy 
$K=\tfrac{1}{2}\sum_i \bm{p}_i^2/m$, the potential energy $\Phi=\tfrac{1}{2}\sum_i m\omega^2\,\bm{r}_i^2$, and their sum $E=K+\Phi$, to later obtain the superstatistical means and variances. 
Finally, the generalized LPV formula, Eq.~\eqref{eq:main}, is then evaluated using the measured values of $\langle E\rangle$, $\langle K\rangle$, and their respective variances, together with $C=3N-1$.

\begin{figure}[h!]
\centering
\includegraphics[width=0.6\textwidth]{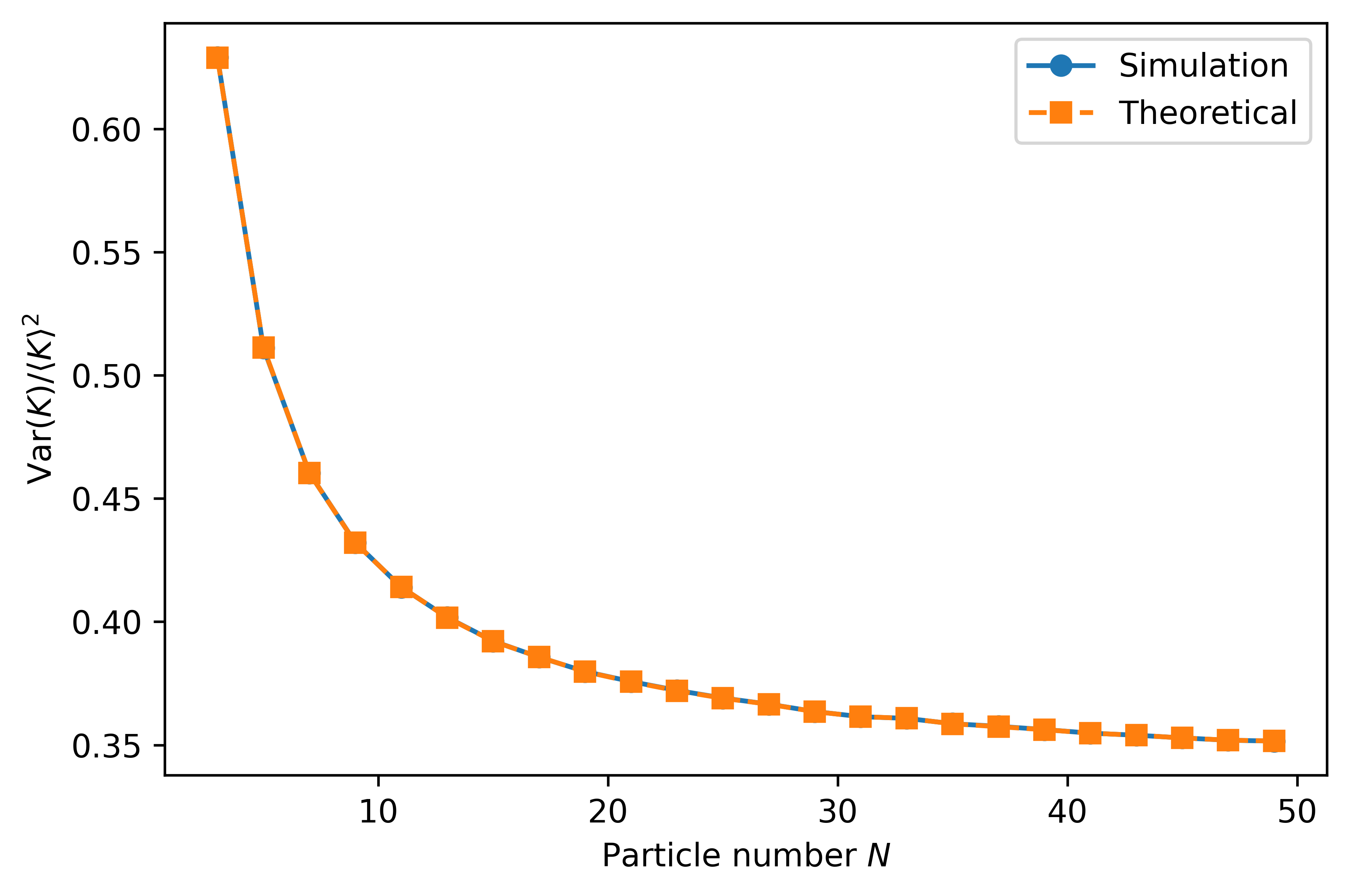}
\caption{Relative variance of the kinetic energy as a function of the number of particles $N$ for a superstatistical ensemble of three--dimensional harmonic oscillators.  Each microstate samples an inverse temperature $\beta$ 
from a gamma distribution with shape $\alpha=5$ and scale $\theta=1$ before drawing coordinates and momenta.  Points correspond to Simulation data and the dashed line is the theoretical prediction from Eq.~\eqref{eq:main}.}
\label{fig:ho-superstat}
\end{figure}

Figure~\ref{fig:ho-superstat} shows the relative variance of the kinetic energy, $\mathrm{Var}(K)/\langle K\rangle^2$, obtained from the superstatistical simulation as a function of the number of particles $N$ ranged 
from 3 to 50, with a sample size of 5$\times10^7$. The values used for scale and shape for the distribution of the $\beta$ values where 1.0 and 5.0 respectively. Numerical results (circles) are compared with the theoretical prediction from Eq.~\eqref{eq:main}, using the measured values of $\langle E\rangle$ and $\mathrm{Var}(E)$.  The agreement is excellent for all system sizes examined, confirming that our generalized LPV relation holds even 
for mixtures of canonical ensembles with fluctuating temperatures.

\subsection{Uniform--energy ensemble}
\label{sec:ue-distribution}

To provide an analytic verification of Eq.~\eqref{eq:main} that does not rely on a particular model or simulation, we now consider a non--equilibrium steady state in which the total energy is not fixed but merely bounded by a cut--off $E_{\max}$. In this ``uniform--energy'' ensemble every phase--space point with total energy $H(\bm R,\bm P)\le E_{\max}$ is sampled with equal probability. Unlike the microcanonical ensemble, where $E$ is fixed, here the energy fluctuates on the interval $[0,E_{\max}]$.

\noindent
Formally, the weight of a microstate is
\begin{equation}
\label{eq:ue_weight}
P(\bm R,\bm P | E_{\max}) = \rho\bigl(H(\bm R,\bm P); E_{\max}\bigr) = \rho_0\Theta\big(E_{\max}-H(\bm R,\bm P)\big), \qquad H(\bm R,\bm P)=K(\bm P)+\Phi(\bm R),
\end{equation}
where $K(\bm P)$ and $\Phi(\bm R)$ denote the kinetic and potential energy, respectively, $\Theta$ is the Heaviside step function, and $\rho_0$ is a normalization constant. Throughout this work we assume a power--law total 
density of states $\Omega(E)\propto E^C$, corresponding to a system with constant microcanonical specific heat $C$, and we denote by $\Omega_K(K;N)$ the $N$-particle kinetic density of states defined in Eq.~\eqref{eq:omegaK}.

\noindent
We are interested in the marginal distribution of kinetic energies,
\begin{equation}
\label{eq:ue_PK_def}
P(K|\params)
=\int_{0}^{\infty}\! \mathrm d\phi\;
\bigl\langle
\delta\big(K(\bm P)-K\big)\,
\delta\big(\Phi(\bm R)-\phi\big)
\bigr\rangle_{\boldsymbol{\lambda}},
\end{equation}
where, following Sec.~\ref{sec:derivation}, we have introduced the value $\phi$ of the potential energy, $\phi\defeq \Phi(\bm R)$. Substituting the uniform weight and using
\begin{equation}
\label{eq:ue_omegaK_id}
\int \mathrm d\bm P\,\delta\big(K(\bm P)-K\big) = \Omega_K(K)
\end{equation}
together with $\mathcal{D}(\phi)\defeq \int \mathrm d\bm R\,\delta\big(\Phi(\bm R)-\phi\big)$, one finds
\begin{equation}
\label{eq:ue_PK_with_step}
P(K|\params) = \rho_0\int_{0}^{\infty}\! \mathrm d\phi\;
\Omega_K(K)\,\mathcal{D}(\phi)\,\Theta\!\bigl(E_{\max}-K-\phi\bigr).
\end{equation}

The step function enforces the constraint $K+\phi\le E_{\max}$, so that for fixed $K$ the potential energy is restricted to $0\le \phi\le E_{\max}-K$. This geometric restriction yields a factorized form
\begin{equation}
\label{eq:PK_uniform_factorized_new}
P(K|\params) = \rho_0\,\Omega_K(K;N)\int_{0}^{E_{\max}-K}\!\mathrm d\phi\,\mathcal{D}(\phi),
\end{equation}
which makes explicit how the available configurational volume decreases as $K$ approaches the cutoff: larger kinetic energies leave less ``room'' for the potential energy.

A key point is that the configurational density of states $\mathcal{D}(\phi)$ is not arbitrary. The total density of states can be written as a convolution,
\begin{equation}
\label{eq:ue_convolution}
\Omega(E) = \int_{0}^{E}\! \mathrm dK\,\Omega_K(K)\,\mathcal{D}(E-K)\propto E^{C},
\end{equation}
and the assumption $\Omega(E)\propto E^C$ fixes the functional form of $\mathcal{D}(\phi)$. As shown within a purely microcanonical framework~\cite{Davis2025}, a power-law total density of states uniquely determines a power-law configurational density of states, yielding
\begin{equation}
\label{eq:d-phi}
\mathcal{D}(\phi)\propto \phi^{\,C-\frac{3N}{2}},
\end{equation}
so that the exponents in the convolution sum to $C$. Inserting Eq.~\eqref{eq:d-phi} into Eq.~\eqref{eq:PK_uniform_factorized_new}, performing the elementary integral and imposing normalization yields
\begin{equation}
\label{eq:ue_PK_beta}
P(K|E_{\max}) = \Gamma(C+1)\frac{K^{\frac{3N}{2}-1}\, \left(E_{\max}-K\right)^{\,C-\frac{3N}{2}}}{(E_{\max})^C\,\Gamma\left(C - \frac{3N}{2} + 1\right)\Gamma\left(\frac{3N}{2}\right)},
\end{equation}
which is a beta distribution for the scaled variable $K/E_{\max}$ with shape parameters
\begin{equation}
\label{eq:ue_beta_params}
\alpha=\frac{3N}{2},
\qquad
\beta=C-\frac{3N}{2}+1.
\end{equation}

\noindent
This provides a closed--form expressions for the mean and variance of $K$,
\begin{equation}
\label{eq:Kmoments_uniform}
\langle K\rangle_{E_{\max}} =\frac{\alpha}{\alpha+\beta}\,E_{\max} = \frac{3N E_{\max}}{2(C+1)},
\qquad
\frac{\langle(\delta K)^2\rangle_{E_{\max}}}{\langle K\rangle_{E_{\max}}^2} = \frac{\beta}{\alpha(\alpha+\beta+1)} = \frac{2C+2-3N}{3N\,(C+2)}.
\end{equation}

Additionally, these expressions do not introduce any new assumptions beyond the power--law form of $\Omega(E)$ and thus constitute an independent check of our generalized LPV relation.

\noindent
The distribution of the total energy follows directly from the uniform weight. Since 
\begin{equation}
\rho(E; E_{\max}) = \begin{cases}
\rho_0\,\,\text{for}\,\,0 \leq E \leq E_{\max}, \\[10pt]
0\,\,\text{otherwise},
\end{cases}
\end{equation}
%
%
combining it with $\Omega(E)\propto E^C$ yields a power law distribution on $[0,E_{\max}]$, 
\begin{equation}
\label{eq:uniform_PE_new}
P(E|E_{\max}) = \Theta(E-E_{\max})\frac{C+1}{E_{\max}^{C+1}}\,E^{C},
\end{equation}
from which the steady--state mean and variance of $E$ are obtained as
\begin{equation}
\label{eq:Erelvar_new}
\langle E\rangle_{E_{\max}} = \frac{C+1}{C+2}\,E_{\max},
\qquad
\frac{\langle(\delta E)^2\rangle_{E_{\max}}}{\langle E\rangle_{E_{\max}}^2} = \frac{1}{(C+1)(C+3)}.
\end{equation}

Finally, inserting Eq.~\eqref{eq:Erelvar_new} into our general expression~\eqref{eq:main} for $\langle(\delta K)^2\rangle_{\boldsymbol{\lambda}}/\langle K\rangle_{\boldsymbol{\lambda}}^2$ and carrying out the algebra shows 
that it reduces to the simple form in Eq.~\eqref{eq:Kmoments_uniform}. In other words, the generalized LPV relation holds exactly for the beta--distribution above. The uniform--energy ensemble therefore serves as a 
transparent analytic test of Eq.~\eqref{eq:main}: no approximations or additional constraints are needed, and the scaling of $\mathcal{D}(\phi)$ is fully determined by the requirement $\Omega(E)\propto E^C$.

Furthermore, this uniform--energy ensemble is an example of a subcanonical state, as we will show below. First, we compute the mean inverse temperature
\begin{equation}
\label{eq:betaS_unif}
\big<\beta_\Omega\big>_{E_{\max}} = \left<\frac{C}{E}\right>_{E_{\max}} = \frac{C+1}{E_{\max}},
\end{equation}
and the variance of the microcanonical inverse temperature,
\begin{equation}
\label{eq:varbetaOm_unif}
\big<(\delta \beta_\Omega)^2\big>_{E_{\max}} = \frac{C+1}{(C-1)(E_{\max})^2}.
\end{equation}

\noindent
Replacing \eqref{eq:betaS_unif} and \eqref{eq:varbetaOm_unif} into \eqref{eq:U}, we see that
\begin{equation}
\U = \big<(\delta \beta_\Omega)^2\big>_{E_{\max}} + \big<{\beta_\Omega}'\big>_{E_{\max}} = -\frac{C+1}{(E_{\max})^2},
\end{equation}
which is always negative, given that we have imposed $C > -1$.

\section{Concluding remarks}

We have generalized the Lebowitz--Percus--Verlet (LPV) formula beyond the microcanonical ensemble, extending its applicability to a broad class of generalized statistical ensembles. This framework clarifies the quantitative 
relationship between kinetic energy fluctuations and the specific heat, and is supported by both analytical derivations and numerical simulation, including the superstatistical harmonic oscillator and the uniform--energy 
ensemble examples presented above. The robustness of the generalized LPV relation underscores the relevance of energy fluctuations as a diagnostic tool across different statistical ensembles.

Although the classification of nonequilibrium steady states into supercanonical and subcanonical categories~\cite{Davis2022b} shows that not all steady states can be described within the superstatistical framework, 
our generalized LPV relation still holds, as its derivation does not rely on the validity of superstatistics. It thus provides a useful tool for characterizing energy fluctuations even in subcanonical states, such as the 
uniform--energy ensemble, where temperature fluctuations cannot be encoded by a probability density $P(\beta|\params)$ under the superstatistical framework.

Future work may extend the present framework beyond the condition of constant microcanonical specific heat, and explore connections with experimental realizations, such as small clusters or driven condensed--matter systems.

\subsection*{Author Contributions}

S. Davis: Conceptualization (equal); Formal analysis (equal); Investigation (equal); Methodology (equal); Writing original draft (equal); Writing -- review \& editing (equal). C. Ruiz: Computational Simulations (equal). C. Loyola: Investigation (equal); Methodology (equal); Validation (equal); Writing original draft (equal); Writing -- review \& editing (equal). C. Femenías: Computational Simulations (equal); J. Peralta: Investigation (equal); Methodology (equal); Writing – original draft (equal); Writing -- review \& editing (equal).
\section*{Acknowledgments}

\noindent
S. D.  and J. P. gratefully acknowledge funding from ANID FONDECYT 1220651 grant. C. R. acknowledges to ANID DOCTORADO 21222208 grant. This research was supported by the supercomputing infrastructures of the NLHPC (CCSS210001) and FENIX (UNAB).

\newpage
\section*{References}

\bibliography{kef}

\end{document}